\documentclass[cits]{PoS}

\title{Simulating acceleration and radiation processes in X-ray binaries}

\ShortTitle{Simulations of radiation and acceleration processes in XRB}

\author{\speaker{R. Belmont}\\
        Centre d'Etude Spatiale des Rayonnements (OMP; UPS; CNRS), 9 avenue du Colonel Roche, BP44346, 31028, Toulouse Cedex 4, France \\
        E-mail: \email{belmont@cesr.fr}}
\author{J. Malzac\\  Centre d'Etude Spatiale des Rayonnements (OMP; UPS; CNRS), 9 avenue du Colonel Roche, BP44346, 31028, Toulouse Cedex 4, France}
\author{A. Marcowith\\  Laboratoire de Physique Th\'eorique et d'Astroparticules, IN2P3/CNRS, Universit\'e MontpellierII, CC 70, place Eug\`ene Bataillon, F-34095 Montpellier Cedex 5, France}

\abstract{The high energy emission of microquasars is thought to originate from high energy particles. Depending on the spectral state, the distribution of these particles can be thermal with a high temperature (typically 100 keV) or non-thermal and extending to even higher energy. The properties of high energy plasmas are governed by a rich microphysics involving particle-particle collisions and particles-photons interactions. 

We present a new code developed to address the evolution of relativistic plasmas. This one-zone code focuses on the microphysics and solves the coupled kinetic equations for particles and photons, including Compton scattering, synchrotron emission and absorption, pair production and annihilation, bremsstrahlung emission and absorption, Coulomb interactions, and prescriptions for additional particle acceleration and heating. It can in particular describe mechanisms such a thermalisation by synchrotron self-absorption and Coulomb collisions.

Using the code, we investigate whether various acceleration processes, namely thermal heating, non-thermal acceleration and stochastic acceleration, can reproduce the different spectral states of microquasars. Premilinary results are presented.  
}

\FullConference{VII Microquasar Workshop: Microquasars and Beyond\\
		 September 1-5 2008\\
		 Foca, Izmir, Turkey}

\begin{document}

\section{Introduction}
The physics of relativistic plasmas is a very general issue. High energy particles are expected to produce the hard emission of all high energy sources. These sources cover a large variety of scales and objects, such as X-ray binaries, AGN, or $\gamma$-ray bursts... Moreover, objects such as microquasars involve different states, with possibly several geometries. Different models have been proposed to account for the high-energy emitting medium, such as the two-temperature flows models, hot coronae, base of the jet, jet emitting disk... Despite this variety, the microphysics at work in the innermost regions of all these high energy sources is expected to be the same.

The physics of high energy plasmas is complex. It results from the combined effects of magnetic fields, the kinetic behaviour of particles and radiation processes. The evolution of such systems is highly non-linear and is described by intregro-differential equations rather than usual differential equations. The cross sections of most processes are quite complicated. And last, Coulomb collisions are often not efficient enough to keep the distributions thermal and to allow for the hydrodynamical or MHD approximations. This complexity makes analytical studies a real challenge and rather points towards the use of the numerical tool. 

Numerical simulations of relativistic plasmas have been performed for more than 20 years now. Two main techniques have been used: the Monte Carlo method (hereafter MC, Pozdnyakov et al., 1980, Stern et al., 1995) and the kinetic approach. The Monte Carlo technique is very efficient in simulating steady state, 3D problems in a given geometry. However, whereas a precise description of geometrical effects can bring interesting constrains, the geometry of the sources is not known and MC simulations are very time consuming, limiting their use to a few long runs. When comparisons with observations, data fitting and exploration of a large parameter space is wanted, kinetic codes are more appropriate. These code can abandon the precise description of the geometry and focus on other features such as radiation processes. The first kinetic codes assumed pure thermal distributions (Fabian et al. 1986, Ghisellini 1988). Most recent simulations accounted for hybrid distributions of particles constituted of a low-energy, thermal component and a higher energy power-law tail (e.g. Lightman \& Zdziarski 1987, Coppi, 1992). However, they included only a limited number of processes, depending on the application they were dedicated to. Moreover such codes cannot address precisely issues such as thermalization mechanisms.

We have developed a new kinetic code that overcomes these limitations by including an exhaustive number of radiation and kinetic processes and by using arbitrary distributions of particles. In section 2, we present the code, its main properties and an illustration of the code capabilities about thermalization by synchrotron self-absorption. In section 3, we present preliminary results on the properties of acceleration mechanisms in X-ray binaries. This work is aimed to discriminate between the different acceleration processes proposed so far (collisions with hot protons, reconnection, shocks, turbulence...). 

\section{The code}
\subsection{Basics}
The code developed is a one-zone code. It does not address the sources geometry explicitly. Instead, we assume an homogeneous sphere of fully magnetized plasma. The properties of this system are fully described by the particle and photon distributions inside this sphere. The exact radiation transfer is not solved throughout the sphere, but the emitted spectrum is computed using the usual {\it escape formalism}, where photons are assumed to have a given, uniform escape-probability. The exact escape probability is chosen to reproduce the steady state results of radiation transfer problems in spherical geometry (Lightman \& Zdziarski 1987, Coppi, 1992, Stern et al. 1995). Also, depending on the precise geometry of the source, the simulated plasma can be illuminated by a flux of external photons. The code accounts for such an illumination by injecting, in the entire sphere, photons of given temperature.

The code is time dependent. It evolves with time three distributions of species: the positron, electron, and photon distributions. Distributions are assumed to be isotropic, so that they only depend on one variable, namely the particle or photon energy. Equations on 1-dimension distributions are very rapid to solve and allow for an efficient code. In contrast to most previous works (see references here before), we do not assume  thermal nor power-law particle distributions. The code allows for arbitrary energy distributions. In addition to the mentioned species, a thermal distribution of protons is described, the temperature of which can also be evolved with time. 

Distributions are discretized in energy bins and the evolution equations are solved bin by bin. The numerical scheme is second order-accurate in space, and first order accurate in time. We use a semi-implicit scheme to overcome the Courant condition limitation. Because the equations are highly non-linear, iterations are made at each time step to converge towards the fully implicit solution. This numerical scheme enables rapid computing and 
robust solutions. 

In addition to a few numerical parameters, the code starts with given initial photon and particles distributions and uses the following set of physical parameters: the system size $R$, the magnetic field intensity $B$, and the luminosity $L_\nu$  and temperature $k_BT_\nu$ of illuminating soft photons. It solves the coupled kinetic equations for photons and particles and evolves their distribution according to the microphysics of relativistic plasmas described in the following section.

\subsection{Microphysics}
Have been included into the code the following processes: Compton scattering, cyclo-synchrotron emission and absorption, pair production and annihilation, lepton-proton bremsstrahlung emission and absorption, lepton-lepton and lepton-proton Coulomb collisions and various prescriptions for additional heating and particle acceleration. \\
$\bullet$ Compton scattering typically up scatters low energy photons by cooling high energy particles. It is described using the full isotropic Klein-Nishina cross section, without approximation (Jones 1968, Nagirner \& Poutanen 1994, Belmont 2008b). This cross section is integrated over the photon and particle distributions. For small-angle scattering, the energy variation in one single scattering event can become smaller than the grid resolution. To prevent severe accuracy issues, we use a Fokker-Planck approximation in this regime. \\
$\bullet$ The cyclo-synchrotron self-absorption is described by the synchrotron emissivity of one single electron in a given magnetic field and the related absorption coefficient. There is no exact formula valid in all regimes for these angle-integrated coefficients. We use a combination of asymptotic expressions for the sub- and ultra relativistic regimes (Ghisellini \& Svensson 1991, Katarzinsky 2006a). The resulting expressions are accurate in all regimes except for very low photon energies where a few harmonics start dominating the spectrum (Marcowith \& Malzac 2003). Synchrotron emissivity and absorption are used to build the coefficients of a Fokker-Planck equation on particles (Ghisellini et al. 1988, 1998).  \\
$\bullet$ Pair production and annihilation are reproduced by using the full isotropic cross sections (Boettcher \& Schlickeiser 1997 and Svensson 1982 respectively), and integrated over the particle and photon distributions. Contrary to Compton scattering, this integration always remains accurate. \\
$\bullet$ Lepton-Proton bremstrahlung self-absorption produces additional soft photons and cools down high energy particles. It is described with the same formalism as synchrotron self-absorption. The full isotropic cross section is used (Heitler 1953, Jauch \& Rohrlich 1976), whose only approximation assumes sub-relativistic protons. For most applications to microquasars however, e-p bremstrahlung is not significant. Electron-electron and electron-positron bremsstrahlung have not been included in the code yet but their effect is though to be also negligible. \\
$\bullet$ Coulomb collisions tend to thermalize the particle distribution by cooling high energy particles and heating low energy ones. In addition, Coulomb interactions with hot protons can heat the lepton populations.  Lepton-lepton and lepton-proton Coulomb collisions are also described with the Fokker-Planck formalism. The Fokker-Planck coefficients are derived from the cross section given in Nayaksin \& Melia 1998. \\
$\bullet$ The code also accounts for particle heating and acceleration. Namely, particles in the code can gain energy by a Coulomb-like heating mechanism;  by stochastic acceleration, or by other non-thermal mechanisms (see section \ref{sec_acc}). When additional acceleration or heating is used, more parameters are required by the code: the heating rate, and some others depending on the included process.

More details on the numerical strategy and the included processes are given in Belmont et al. 2008a. The code developed reproduces most relevant processes of high energy plasmas such as the Comptonizing medium of microquasars. It uses general numerical schemes and cross sections accurate in all regimes. In particular, the photon distribution can span the entire electro-magnetic spectrum, from radio wavelengths (or lower energy), to TeV emission (or higher). The code can deal with particles from the sub-relativistic regime (with momentum $p>10^{-7} m_ec$) to the ultra-relativistic regime ($p<10^{-7} m_ec$). This makes the code very general and enables its use for many different astrophysical applications: X-ray binaries, AGN, $\gamma$-ray bursts...

\subsection{Example: the Synchrotron Boiler}
One of the new features of our code is the synchrotron self-absorption, which can play a crucial role in microqusars. For the first time this process is consistently taken into account in a global code, together with other radiation and kinetic processes. Here we present examples of runs made with our code to illustrate the {\it synchrotron boiler} mechanism (Ghisellini et al. 1988, 1998). 

\begin{figure}[h!]
\includegraphics[width=.5\textwidth]{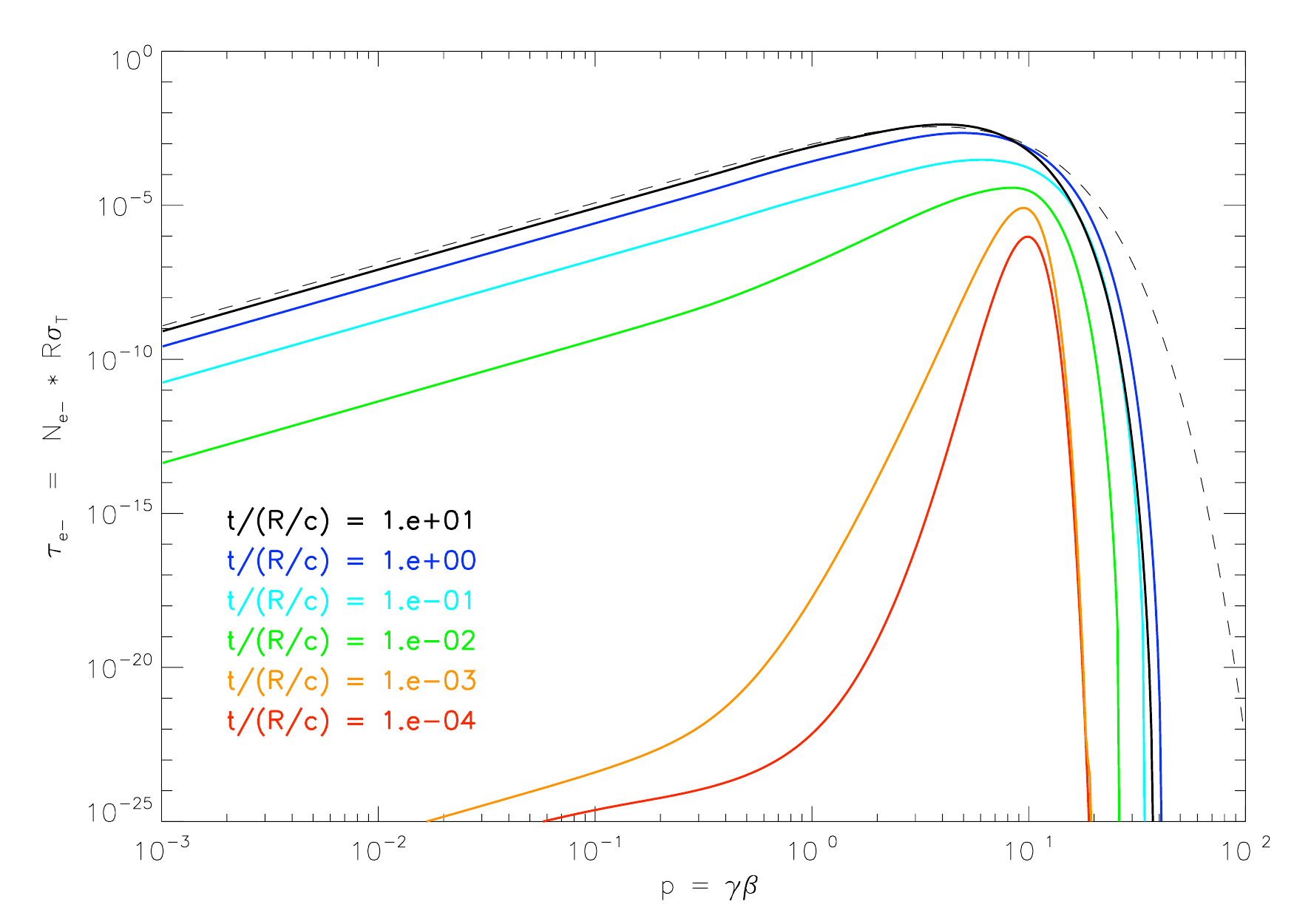} \includegraphics[width=.5\textwidth]{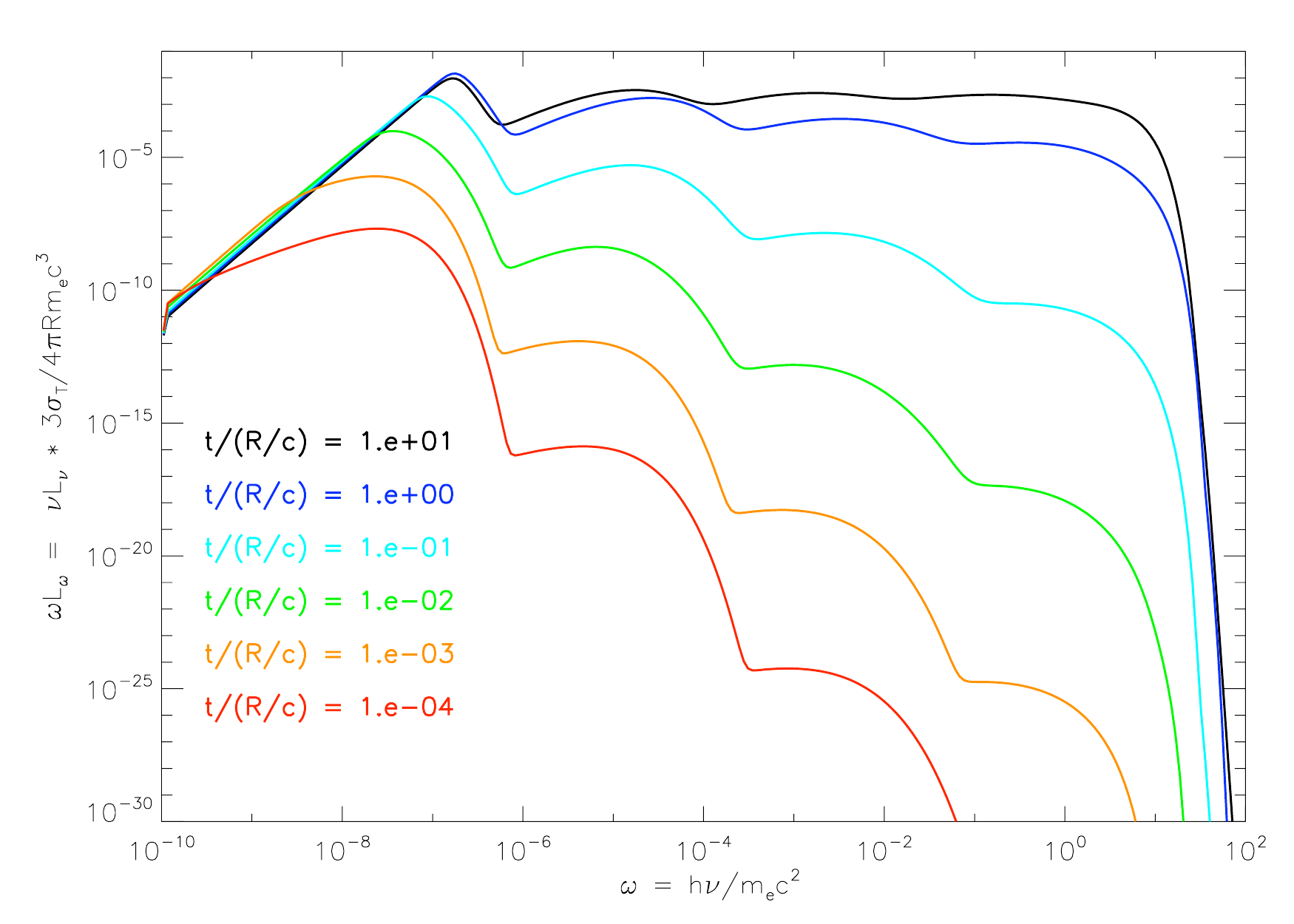} 
\caption{Time evolution of the particle (left panel) and photon (right panel) distributions under effect of synchrotron self-absorption ($B=5.5\times 10^3$ G) and Compton scattering. Particle are injected at $\gamma=10$ with a constant rate ($ \dot{N}_{e-} = 10^7$ s$^{-1}$cm$^{-3}$) and escape in one dynamical time $R/c$. The Maxwell-Boltzmann distribution of same energy is over-plotted in dashed line.The size is here typical of AGN: $R=10^{13}$ cm. } 
\label{sboiler} 
\end{figure}
Fig. \ref{sboiler} shows the time evolution of the particle distribution and the emitted spectrum when only self-absorbed synchrotron radiation and Compton scattering are taken into account. Particles are continuously injected at high energy and are allowed to escape freely. The initial conditions correspond to an empty sphere. As time goes, the number of particles increases until it reaches a steady state, where losses equals the injection.  The injected high energy particles start emitting soft synchrotron radiation. As a result, they cool down and form the low energy part of the distribution. When they reach very low energies, the particles start absorbing photons, which prevents them to cool further down. Simultaneously, high energy particles also up-scatter the soft synchrotron photons, producing the hard tail observed on the right panel. 

In steady state, the particle distribution is quasi-Maxwellian, at least at low energy. A deviation is only observed at high energy, where particles are injected and cool down. This example illustrates the thermalization by synchrotron self-absorption. The efficiency of this mechanism depends on the parameter regime. It is very efficient for optically thin, magnetized sources with weak external illumination. For dense plasmas, Coulomb collisions become the dominant thermalization process, whereas strong Compton cooling prevents particles from being heated by absorption in strongly illuminated sources. In microquasars, thermalization by the synchrotron boiler mechanism can explain various properties of the state transition in X-ray binaries (Malzac \& Belmont 2008, and see also J. Malzac in these proceedings).

\section{Constraining acceleration processes in X-ray binaries}
\label{sec_acc}

Among others interesting issues, the code was used to address the heating problem of microquasars corona.

X-ray binaries exhibit two interesting canonical states: the low-hard state (LS hereafter), characterized by a rather low luminosity and a power-low spectrum extending to high energy with a cut-off at around 100 keV, and the high-soft state (HS), characterized by a higher luminosity and a spectrum composed by a soft excess at about a few keV and a steep high-energy tail extending to MeV energies, with no hint for a cut-off.  Spectra in the LH state are well reproduced by thermal Comptonization models with plasma temperatures of about 100 keV, whereas spectra in the HS state are best reproduced by a black-body emission from the disk and its Comptonisation by high energy, non-thermal particles. 

In all states, high energy particles must cool very quickly because of the various radiation cooling processes such a Compton scattering with soft photons. To keep a significant fraction of emitting high energy particle, power must be continuously injected into this hot medium. Various mechanisms have been proposed to account for this power supply, such as heating by hot protons, magnetic reconnection, turbulent heating, and shocks. However, the coupled effect of acceleration and all relevant radiation processes have not been studied in a consistent approach yet, and the mechanisms at work in microquasars remain poorly constrained. Moreover, the different particle distributions inferred in the different states (thermal in the LS state and non-thermal in the HS state) represent a real challenge to the modelling. 

To investigate the various channels through which the power can be supplied to the Comptonizing medium, three prescriptions have been included in the code. 

\subsection{Thermal heating}
Models such as two-temperature disks assume that protons and electrons are almost decoupled (Narayan \& Yi 1994). Protons are heated by a strong anomalous viscosity. Since they cannot radiate their energy and are only weakly coupled to electrons, they reach very high temperatures (typically 10-100 MeV). The coupling with leptons, although weak, allows to heat electrons, that then radiate this energy. This heating is by nature thermal and has been applied to reproduce X-ray binary spectra in the LH state (e.g. Esin et al. 1997). 

To model such thermal heating, the code consistently describes the Coulomb collisions with a thermal population of protons. To bring more direct constraints, the heating rate, instead of the proton temperature, is set and the proton temperature responsible for this rate is computed consistently with the electron distribution. 

Simulations with the code confirm that Coulomb heating produces thermal distributions and thermal spectra such as those observed in the LH state. Spectra of HS states can not be reproduced by such a pure thermal heating, nor can the non-thermal tails observed in some observations of microquasars in the LS state (McConnell et al. 2002). 

\subsection{Non thermal acceleration}
Particles can also be accelerated by magnetic reconnection or shocks. Reconnection has been proposed to occur in an active corona overlying a turbulent, magnetized accretion disk (Galeev et al. 1979). Because of turbulence and buoyancy, field loops rise above the disk and reconnect into the corona, accelerating particles to the required energies. The detailed microphysics of reconnection and shocks is still uncertain and is out scope of this study. These processes are however thought to produce power-law distributions of particles. 

Non-thermal particle acceleration is mimicked by injecting high energy particles with a power-law distribution. To conserve the total number of particles, leptons are physically removed from the system before they are re-injected as a power-law distribution. This injection is characterized by the minimal and maximal Lorentz factor of accelerated particles $\gamma_{\rm min}$ and $\gamma_{\rm max}$, the slope of the power-law $\Gamma$ and the injection rate.

Results of simulations with non-thermal acceleration are presented in theses proceedings by J. Malzac (see also, Malzac \& Belmont 2008). We just remind here the main results. Our simulations show that the non-thermal HS spectra are well reproduced by assuming non-thermal acceleration. They also show that the synchrotron self-absorption and Coulomb collisions in the LH state are efficient in thermalizing the distribution, and that models with pure non-thermal acceleration can also reproduce thermal spectra. As a result, models with one unique, non-thermal acceleration process can explain both LH and HS states of microquasars, the state transition resulting mostly from a change in the illumination from the cold accretion disk. 

\subsection{Stochastic acceleration}
Last, in turbulent media, particles can also be accelerated by resonant interactions with plasma waves, such as Alf\'en or fast magnetosonic modes (Li \& Miller 1997, Katarsinsky et al. 2006b). Stochastic acceleration results in a diffusive process where some particles are accelerated and others are cooled down. 

Stochastic acceleration is described by the 2nd order Fermi equations. To account for the fact that only particles above some energy are resonant with waves and thus accelerated, we added a threshold in the equations. The precise acceleration rate and threshold energy directly result from the details of the plasma turbulence, which is poorly constrained. Instead, these parameters are varied in the simulations in order to constrain the plasma properties. 
\begin{figure}[h!]
\includegraphics[width=.5\textwidth]{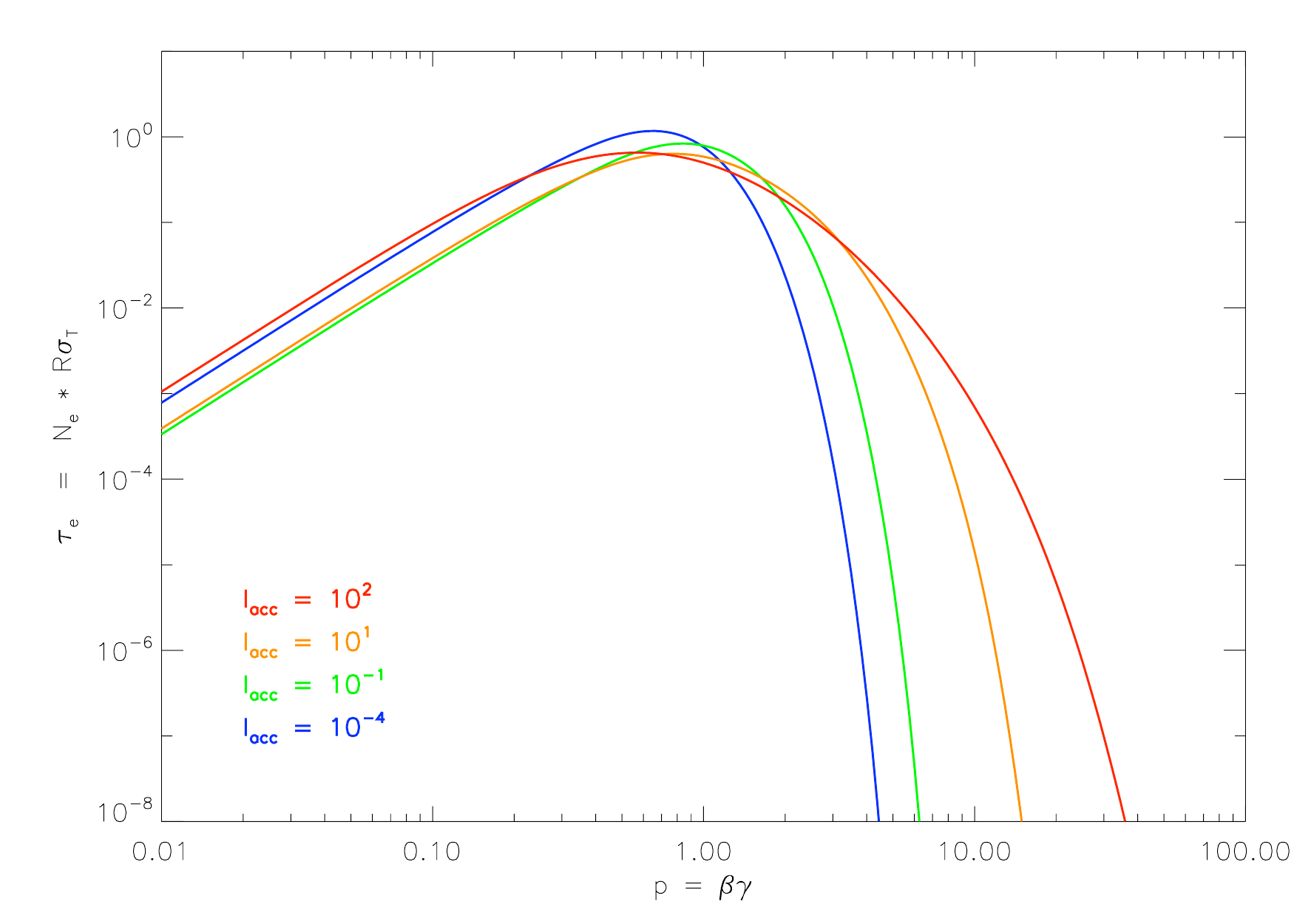} \includegraphics[width=.5\textwidth]{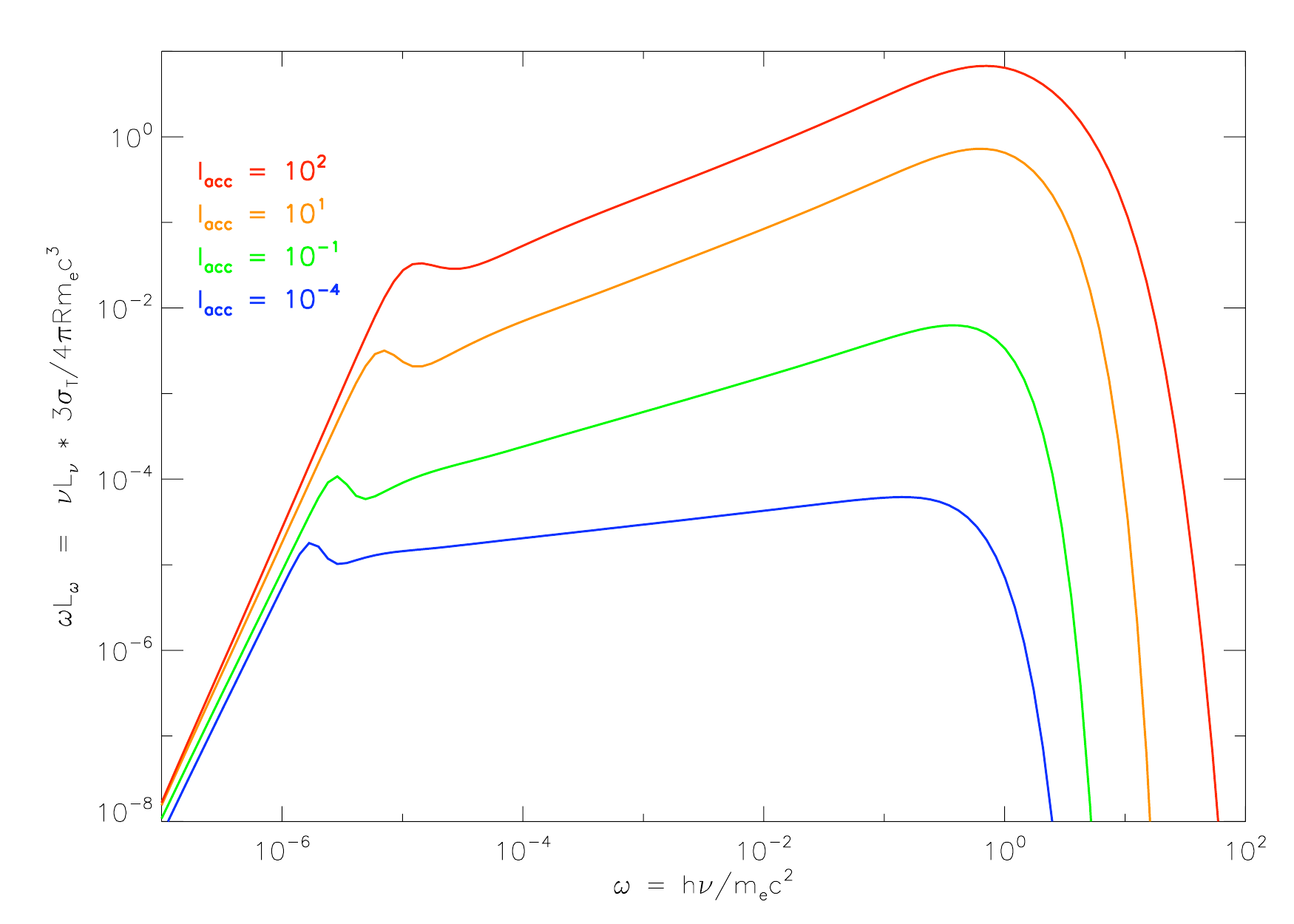} 
\caption{Steady state solutions for particle distributions (left panel) and photon spectra (right panel) for simulations with stochastic acceleration of different acceleration rate and no threshold. Only stochastic acceleration, self-absorbed synchrotron radiation and Compton scattering are switched on. The power injected into particles is $P= 1.8\times 10^{32}, 1.8\times10^{35}, 1.8\times10^{37}$, and $1.8\times10^{38}$ erg~s$^{-1}$, for the blue, green, orange, and red curves respectively. The other parameters are: $R=5\times10^{7}$ cm, $B= 7.9\times10^5$~G, $\tau=1$.} 
\label{acc1} 
\end{figure}
Simulations are completed including only Compton scattering, self-absorbed synchrotron radiation and stochastic acceleration. As no external illumination is added, all soft photons are produces by synchrotron emission. 

Figure \ref{acc1} shows the steady state particle distributions and photon spectra when the acceleration rate in varied, with no threshold. As can be seen, the particle distributions are always quasi-thermal. Second order Fermi acceleration process is a diffusive process and it cannot produce hard tails of particles. The higher deviation from a Maxwellian distribution appears at high energy for very large acceleration rates. The corresponding spectra are typical of thermal Comptonization, as those observed in the LH state of X-ray binaries. 

\begin{figure}[h1]
\includegraphics[width=.5\textwidth]{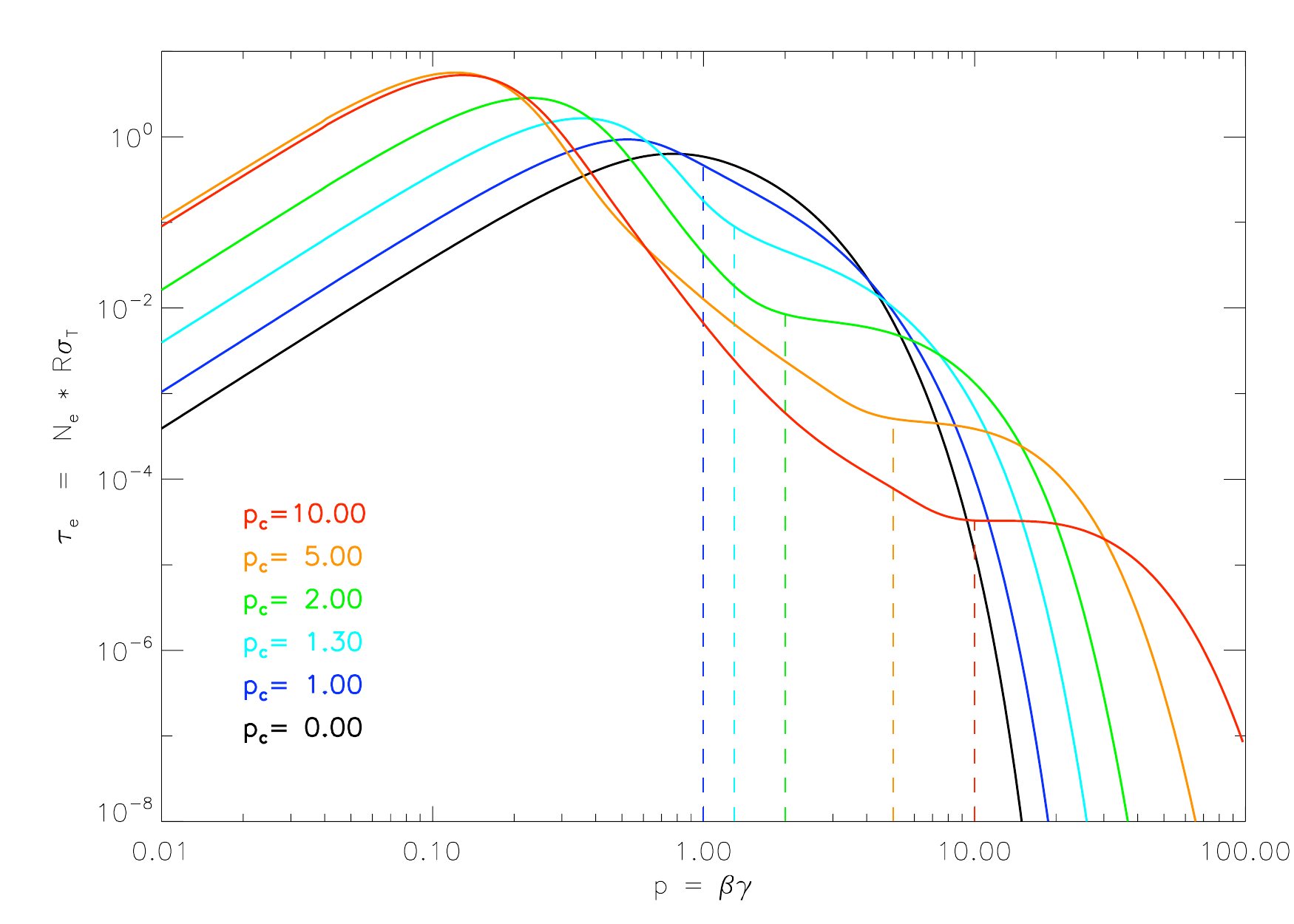} \includegraphics[width=.5\textwidth]{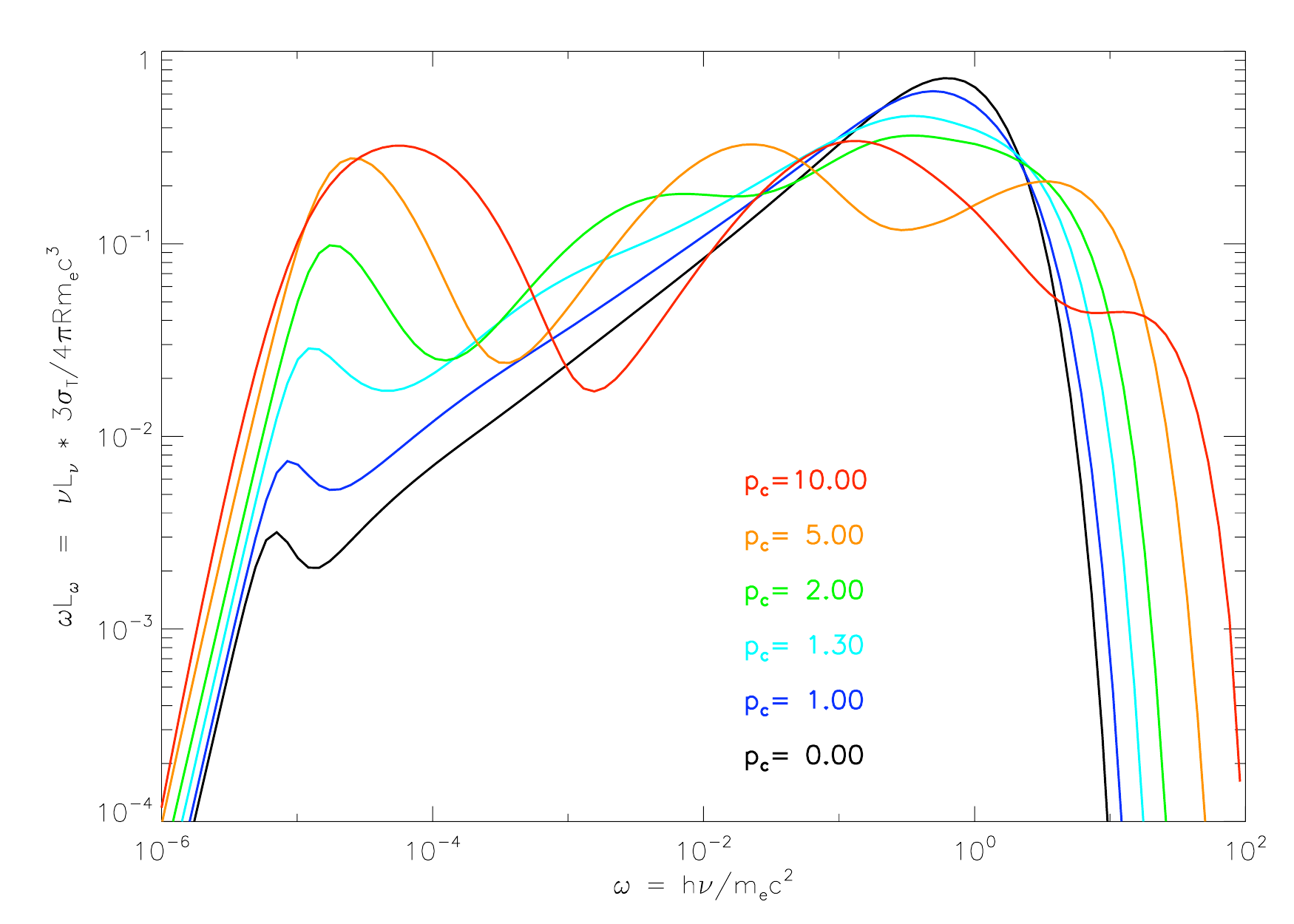} 
\caption{Steady state solutions for particle distributions (left panel) and photon spectra (right panel) for simulations with stochastic acceleration of constant rate ($P=1.8\times10^{37}$~erg~s$^{-1}$) and different energy thresholds. Only stochastic acceleration, self-absorbed synchrotron radiation and Compton scattering are switched on. The threshold Lorentz factors are $\gamma_{\rm min}= 1, 1.4, 1.6, 2.2, 5$, and 10 for the blue, green, orange, and red curves respectively. Other parameters are: $R=5\times10^{7}$ cm, $B=7.9\times10^5$~G, $\tau=1$.} 
\label{acc2} 
\end{figure}
We also varied the threshold energy for acceleration at constant acceleration rate. Results are presented in Fig. \ref{acc2}. For low energy threshold, the distributions are thermal and so are the emitted spectra. However, these results show that when the threshold reaches relativistic energies ($\gamma_{\rm min}  \ge 2$), the distributions strongly deviate from the Maxwell-Boltzmann distribution. Although the low energy part is always thermal, hard tails of particles are observed. The corresponding spectra are also strongly non-thermal. 

These results show that stochastic acceleration is a potential candidate to explain the presence of high energy particles in X-ray binaries. Depending in the threshold energy, it can produce both thermal and non-thermal distributions. Although at this stage, the non-thermal spectra look more like blazar spectra than spectra of microquasars, only Compton scattering was included in these runs and the parameter space has not been investigated yet. Other regimes and other ingredients such as external illumination are expected to modify significantly the spectrum.

\section{Conclusion}
We presented a new one-zone, kinetic code developed to study the properties of the relativistic plasmas in high energy sources. For the first time, our code solved the coupled kinetic equationd for arbitrary distributions of particles and photons, including most relevant processes: Compton scattering, synchrotron and bremsstrahlung self-absorption, pair production/annihilation, Coulomb collisions and prescriptions for thermal, non-thermal and stochastic particle acceleration. This code is very general and can be used to model not only X-ray binaries, but also AGN and $\gamma$-ray bursts.

As a first example of the code capabilities, we presented some preliminary results on a recent work on acceleration processes in the Comptonizing medium of microquasars. We showed qualitatively that 1) thermal heating can account for the thermal spectra in the low-hard state of microquasar, 2) non-thermal acceleration can reproduce both the non-thermal high-soft state emisson and the thermal low hard state spectra, 3) stochastic acceleration usually produces thermal spectra but can also form non thermal tails for relativistic acceleration threshold. Wider exploration of the parameter space and real data fitting are required to give more reliable constraints.


\begin{thebibliography}{99}
\bibitem{Belmont08a} Belmont, R., A\&A , 2008a, A\&A, in press. arXiv:0808.1258
\bibitem{Belmont08b} Belmont, R., A\&A , 2008b, submitted to A\&A
\bibitem{BS97} Boettcher, M. \& Schlickeiser R., 1997
\bibitem{Coppi92} Coppi, P.~S.\ 1992, MNRAS, 258, 657
\bibitem{1997ApJ...489..865E} Esin, A.~A., McClintock, J.~E., \& Narayan, R.\ 1997, ApJ, 489, 865
\bibitem{FBGPC86} Fabian, A.~C., Guilbert, P.~W., Blandford, R.~D., Phinney, E.~S., \& Cuellar, L.\ 1986, MNRAS, 221,
931
\bibitem{Ghisellini87} Ghisellini, G.\ 1987, MNRAS, 224, 1 
\bibitem{GGS88} Ghisellini, G., Guilbert, P.~W., \& Svensson, R.\ 1988, ApJl, 334, L5
\bibitem{GS91} Ghisellini, G., \& Svensson, R.\ 1991, MNRAS, 252, 313
\bibitem{GHS98} Ghisellini, G., Haardt, F., \& Svensson, R.\ 1998, MNRAS, 297, 348
\bibitem{Ghisellini98} Ghisellini, G., Celotti, A., Fossati, G., Maraschi, L., \& Comastri, A.\ 1998, MNRAS, 301, 451
\bibitem{Heitler54} Heitler, W.\ 1954, International Series of Monographs on Physics, Oxford: Clarendon, 1954, 3rd ed.
\bibitem{JR76} Jauch, J.~M., \& Rohrlich, F.\ 1976, Texts and Monographs in Physics, New York: Springer, 1976, 2nd ed.
\bibitem{Jones68} Jones, F.~C.\ 1968, Physical Review , 167, 1159
\bibitem{KGSG06} Katarzy{\'n}ski, K., Ghisellini, G., Svensson, R., \& Gracia, J.\ 2006a, A\&A, 451, 739
\bibitem{Katar06} Katarzy{\'n}ski, K., Ghisellini, G., Mastichiadis, A., Tavecchio, F., \& Maraschi, L.\ 2006b, A\&A, 453, 47 
\bibitem{Li97} Li, H., \& Miller, J.~A.\ 1997, ApJl, 478, L67 
\bibitem{LZ87} Lightman, A.~P., \& Zdziarski, A.~A.\ 1987, ApJ, 319, 643
\bibitem{mj00} Malzac, J., \& Jourdain, E.\ 2000, A\&A, 359, 843 
\bibitem{Marcowith03} Marcowith, A., \& Malzac, J.\ 2003, A\&A, 409, 9
\bibitem{2002ApJ...572..984M} McConnell, M.~L., et  al.\ 2002, ApJ, 572, 984
\bibitem{NP94} Nagirner, D.~I., \& Poutanen, J.\ 1994, Single Compton scattering, Astrophysics and Space Physics Reviews,   vol.~9, part 1.~ Amsterdam: Harwood Academic Publishers,  c1994, 83 pages.,
\bibitem{NY94} Narayan, R., \& Yi, I.\ 1994, ApJL, 428, L13
\bibitem{NM98} Nayakshin, S., \& Melia, F.\ 1998, ApJs, 114, 269
\bibitem{PSS80} Pozdnyakov, L.~A., Sobol, I.~M., \& Syunyaev, R.~A.\ 1980, Comptomization and radiation spectra of X-ray sources.~ Calculation of the Monte Carlo method,  Rept.~Pr-447 Acad.~of Sci.~USSR, Moscow, 1978  12 p,
\bibitem{Stern95} Stern, B.~E., Begelman, M.~C., Sikora, M., \& Svensson, R.\ 1995, MNRAS, 272, 291
\bibitem{Svensson82} Svensson, R.\ 1982, ApJ, 258, 321

\end{thebibliography}
\end{document}